\documentclass[12pt]{article}
\usepackage{axodraw}

\catcode`@=12
\begin{document}
\thispagestyle{empty}
\begin{flushright} 
UCRHEP-T368\\ 
January 2004\
\end{flushright}
\vspace{0.5in}
\begin{center}
{\LARGE	\bf Lepton Family Symmetry\\ and Neutrino Mass Matrix\\}
\vspace{1.5in}
{\bf Ernest Ma\\}
\vspace{0.2in}
{\sl Physics Department, University of California, Riverside, 
California 92521\\}
\vspace{1.5in}
\end{center}
\begin{abstract}\
The standard model of leptons is extended to accommodate a discrete 
$Z_3 \times Z_2$ family symmetry.  After rotating the charged-lepton mass 
matrix to its diagonal form, the neutrino mass matrix reveals itself as 
very suitable for explaining atmospheric and solar neutrino oscillation 
data.  A generic requirement of this approach is the appearance of three 
Higgs doublets at the electroweak scale, with observable flavor violating 
decays.
\end{abstract}
\newpage
\baselineskip 24pt

In the standard model of particle interactions, the fermion mass matrices are 
{\it a priori} completely arbitrary, and yet data seem to indicate specific 
patterns which are not yet understood theoretically.  Recent experimental 
advances in measuring the neutrino-oscillation parameters in atmospheric 
\cite{atm} and solar \cite{sol} data have now fixed the $3 \times 3$ lepton 
mixing matrix to a large extent. This information is not sufficient to fix 
all the elements of the neutrino mass matrix ${\cal M}_\nu$ (assumed here 
to be Majorana at the outset), but is enough to fix its approximate 
\underline {form} in terms of a small number of parameters \cite{allp}.  
However, this works only in the basis $(\nu_e, \nu_\mu, \nu_\tau)$, i.e. 
where the charged-lepton mass matrix ${\cal M}_l$ has been assumed diagonal.

If a symmetry is behind the observed pattern of ${\cal M}_\nu$, then it must 
also apply to ${\cal M}_l$.  The realization of this in a \underline 
{complete} theory of leptons has only been done in a few cases.  Some recent 
examples are those based on the symmetries of geometric objects, i.e. $S_3$ 
(equilateral triangle) \cite{s3}, $D_4$ (square) \cite{d4}, and $A_4$ 
(tetrahedron) \cite{a4,bmv}. In this paper, a much simpler and more flexible 
model is proposed, based on the discrete symmetry $Z_3 \times Z_2$.  
[Because of the choice of representations, this turns out to be equivalent 
to a specific realization of $S_3$.]

The key idea which allows ${\cal M}_\nu$ to be more restricted than 
${\cal M}_l$ is that they come from different mechanisms.  Whereas the 
former comes from the naturally small vacuum expectation value ($vev$) of 
a single heavy Higgs triplet $\xi = (\xi^{++}, \xi^+, \xi^0)$ \cite{masa98}, 
thus dispensing with the usual three heavy singlet neutrinos, the latter comes 
from the $vev$'s of \underline {three} Higgs doublets \cite{z4,z3}.  As shown 
below, the use of $Z_3 \times Z_2$ as a lepton family symmetry results in 2 
parameters for ${\cal M}_\nu$ and 5 parameters for ${\cal M}_l 
{\cal M}_l^\dagger$.  After rotating ${\cal M}_l {\cal M}_l^\dagger$ to its 
diagonal form, i.e. $(m_e^2, m_\mu^2, m_\tau^2)$, then under a condition to 
be derived below, the neutrino mass matrix will become \cite{z3}
\begin{equation}
{\cal M}_\nu = \pmatrix {A-B & 0 & 0 \cr 0 & A & -B \cr 0 & -B & A},
\end{equation}
which is very suitable as a starting point for explaining atmospheric and 
solar neutrino oscillations. The addition of a charged Higgs singlet $\chi^+$ 
and the soft breaking of $Z_3 \times Z_2$ will then lead to a complete 
satisfactory description of all data, including the possibility of $CP$ 
violation, i.e a nonzero complex $U_{e3}$.

The representations of $Z_3$ are denoted by 1, $\omega$, $\omega^2$, where 
\begin{equation}
\omega = e^{2 \pi i/3} = -{1 \over 2} + i \sqrt {3 \over 2},
\end{equation}
with $1 + \omega + \omega^2 = 0$.  Let the standard model be augmented with 
a Higgs triplet $\xi = (\xi^{++}, \xi^+, \xi^0)$, three doublets $\phi_i = 
(\phi_i^0, \phi_i^-)$ and a charged singlet $\chi^+$, in addition to the 
usual three lepton doublets $(\nu_i, l_i)$ and three singlets $l_i^c$. 
Under $Z_3$, let
\begin{eqnarray}
&& l_1, ~l_1^c, ~\phi_1, ~\xi, ~\chi \sim 1, \\ 
&& l_2, ~l_2^c, ~\phi_2 \sim \omega, \\ 
&& l_3, ~l_3^c, ~\phi_3 \sim \omega^2.
\end{eqnarray}
Under $Z_2$, let
\begin{equation}
l_2 \leftrightarrow l_3, ~~l_2^c \leftrightarrow l_3^c, ~~\phi_2 
\leftrightarrow \phi_3, ~~\chi \leftrightarrow -\chi.
\end{equation}
Then the Yukawa couplings of
\begin{equation}
\nu_i \nu_j \xi^0 - \left( {\nu_i l_j + l_i \nu_j \over \sqrt 2} \right) \xi^+ 
+ l_i l_j \xi^{++}
\end{equation}
imply that ${\cal M}_\nu$ is of the form \cite{ma03}
\begin{equation}
{\cal M}_\nu = \pmatrix {a & 0 & 0 \cr 0 & 0 & b \cr 0 & b & 0}.
\end{equation}
Note that ${\cal M}_\nu$ is proportional to $\langle \xi^0 \rangle$ which \
can be naturally small if $m_\xi^2$ is positive and large.  On the other 
hand, the Yukawa couplings of
\begin{equation}
l^c_i (l_j \phi_k^0 - \nu_j \phi_k^-)
\end{equation}
imply that the mass matrix linking $l$ to $l^c$ is given by
\begin{equation}
{\cal M}_l = \pmatrix {c & f & f \cr g & d & e \cr g & e & d},
\end{equation}
where $v_2 = v_3$ has been assumed for $\langle \phi_k^0 \rangle$.  [This 
assumption will be relaxed later to accommodate the soft breaking of $Z_2$. 
The role of $\chi^+$ will also be explained.]

To rotate ${\cal M}_l$ to its diagonal form, i.e.
\begin{equation}
V_L {\cal M}_l V_R^\dagger = \pmatrix {m_e & 0 & 0 \cr 0 & m_\mu & 0 \cr 
0 & 0 & m_\tau},
\end{equation}
consider
\begin{equation}
{\cal M}_l {\cal M}_l^\dagger = \pmatrix {C & F & F \cr F^* & D & E \cr F^* & 
E & D},
\end{equation}
where
\begin{eqnarray}
C &=& |c|^2 + 2|f|^2, \\ 
D &=& |d|^2 + |e|^2 + |g|^2, \\ 
E &=& d e^* + e d^* + |g|^2, \\ 
F &=& c g^* + f (d^* + e^*).
\end{eqnarray}
Then
\begin{equation}
V_L {\cal M}_l {\cal M}_l^\dagger V_L^\dagger = \pmatrix {m_e^2 & 0 & 0 \cr 
0 & m_\mu^2 & 0 \cr 0 & 0 & m_\tau^2}
\end{equation}
and the neutrino mass matrix in the basis $(\nu_e, \nu_\mu, \nu_\tau)$ is 
given by
\begin{equation}
{\cal M}_\nu = V_L \pmatrix {a & 0 & 0 \cr 0 & 0 & b \cr 0 & b & 0} V_L^T.
\end{equation}
As a trial, let
\begin{equation}
V_L = \pmatrix {0 & -i/\sqrt 2 & i/\sqrt 2 \cr -1/\sqrt 2 & 1/2 & 1/2 \cr 
1/\sqrt 2 & 1/2 & 1/2},
\end{equation}
then $V_L {\cal M}_l {\cal M}_l^\dagger V_L^\dagger$ becomes
\begin{equation}
\pmatrix {D-E & 0 & 0 \cr 0 & (C+D+E)/2 - \sqrt 2 ReF & (D+E-C)/2 - i\sqrt 2 
ImF \cr 0 & (D+E-C)/2 + i\sqrt 2 ImF & (C+D+E)/2 + \sqrt 2 ReF}.
\end{equation}
Comparing this with Eq.~(17), it is clear that under the condition
\begin{equation}
C = D+E, ~~~ ImF = 0,
\end{equation}
the charged-lepton matrix is diagonalized by the $V_L$ of Eq.~(19) with
\begin{eqnarray}
m_e^2 &=& D-E, \\ 
m_\mu^2 &=& (C+D+E)/2 - \sqrt 2 ReF, \\ 
m_\tau^2 &=& (C+D+E)/2 + \sqrt 2 ReF,
\end{eqnarray}
and in this basis, the neutrino mass matrix of Eq.~(18) is given by
\begin{equation}
{\cal M}_\nu = \pmatrix {b & 0 & 0 \cr 0 & (b+a)/2 & (b-a)/2 \cr 
0 & (b-a)/2 & (b+a)/2},
\end{equation}
which is identical to Eq.~(1) with the substitution $b=A-B$ and $a=A+B$. 
The eigenvalues and eigenvectors of ${\cal M}_\nu$ are then
\begin{eqnarray}
b &:& \nu_e, \\ 
b &:& (\nu_\mu+\nu_\tau)/\sqrt 2, \\ 
a &:& (-\nu_\mu+\nu_\tau)/\sqrt 2.
\end{eqnarray}
This explains atmospheric neutrino oscillations with
\begin{equation}
\Delta m^2_{atm} = a^2-b^2, ~~~ \sin^2 2 \theta_{atm} = 1,
\end{equation}
but $\Delta m^2_{sol} = 0$, which is nevertheless a good first approximation. 
To split the two degenerate neutrino masses responsible for solar neutrino 
oscillations, consider the Yukawa coupling
\begin{equation}
(\nu_2 l_3 - l_2 \nu_3) \chi^+ = -i[\nu_e (\mu + \tau)/\sqrt 2 - e (\nu_\mu 
+ \nu_\tau)/\sqrt 2],
\end{equation}
which is the only one allowed under $Z_3 \times Z_2$.  In addition, the 
most general trilinear scalar couplings
\begin{equation}
\chi^+ (\phi_i^0 \phi_j^- - \phi_i^- \phi_j^0)
\end{equation}
are assumed, thus breaking $Z_3 \times Z_2$ but only softly.  As a result, 
there are one-loop contributions to ${\cal M}_\nu$ as shown in Fig.~1, and 
three nonzero parameters will emerge, i.e. $\Delta m^2_{sol}$, 
$\tan^2 \theta_{sol}$, and $U_{e3}$.

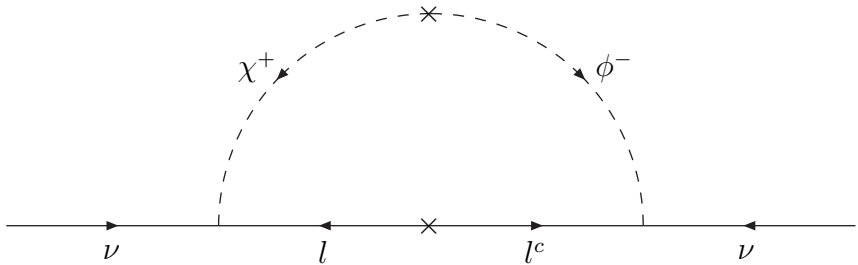
\begin{figure}[htb]
\begin{center}
\begin{picture}(360,120)(0,0)
\ArrowLine(20,10)(100,10)
\ArrowLine(180,10)(100,10)
\ArrowLine(180,10)(260,10)
\ArrowLine(340,10)(260,10)
\DashArrowArc(180,10)(80,90,180){4}
\DashArrowArcn(180,10)(80,90,0){4}
\Text(60,0)[]{$\nu$}
\Text(300,0)[]{$\nu$}
\Text(140,0)[]{$l$}
\Text(220,0)[]{$l^c$}
\Text(180,10)[]{$\times$}
\Text(180,90)[]{$\times$}
\Text(115,70)[]{$\chi^+$}
\Text(250,70)[]{$\phi^-$}

\end{picture}
\end{center}
\caption{One-loop contributions to the neutrino mass matrix.}
\end{figure}

In the original $\nu_{1,2,3}$ basis, i.e. that of Eq.~(8), the radiative 
corrections of Fig.~1 have the form
\begin{equation}
{\cal M}_{rad} = \pmatrix {0 & r & s \cr r & p & t \cr s & t & q}.
\end{equation}
Rotating to the basis $\{\nu_e, (\nu_\mu+\nu_\tau)/\sqrt 2, (-\nu_\mu+\nu_\tau)
/\sqrt 2\}$ and using Eqs.~(26) to (28), the neutrino mass matrix is then 
given by
\begin{equation}
{\cal M}_\nu = \pmatrix {b+t-(p+q)/2 & i(q-p)/2 & i(s-r)/\sqrt 2 \cr 
i(q-p)/2 & b+t+(p+q)/2 & (r+s)/\sqrt 2 \cr i(s-r)/\sqrt 2 & (r+s)/\sqrt 2 
& a}.
\end{equation}
The two-fold degeneracy of the solar neutrino doublet is now lifted with
\begin{equation}
\Delta m^2_{sol} = 4(b+t)\sqrt {pq},
\end{equation}
and
\begin{equation}
\tan^2 \theta_{sol} = \left( {1-x \over 1+x} \right)^2, ~~~ 
x = \sqrt {q \over p}.
\end{equation}
For $x \simeq 0.22$, $\tan^2 \theta_{sol} \simeq 0.41$, as indicated 
\cite{fit} by most recent data.  A nonzero
\begin{equation}
U_{e3} \simeq {i(r-s) \over \sqrt 2 (a-b-t)}
\end{equation}
is also obtained.  Note that the phase of $U_{e3}$ is undetermined because 
$r-s$ is in general complex.  The soft terms of Eq.~(31) break $Z_2$, hence 
the condition $v_2 = v_3$ assumed previously should also be relaxed, in 
which case the zero entries of Eq.~(20) would become nonzero. This means 
another contribution to $U_{e3}$.  The condition of Eq.~(21) may also be 
relaxed to account for any possible deviation from maximum mixing in 
atmospheric neutrino data.

The representation content of lepton families and Higgs multiplets under 
$Z_3 \times Z_2$ proposed here is such that the resulting model is equivalent 
to the following specific realization of $S_3$.  There are 3 irreducible 
representations of $S_3$: \underline {1}$^+$, \underline {1}$^-$, 
\underline {2}.  If \underline {1}$^+$ and \underline {2} are chosen as 
representations here with \underline {1}$^+$ as 1 and \underline {2} as 
$(\omega,\omega^2)$ under $Z_3$, together with $Z_2$ of Eq.~(6), then the 
group multiplication rules of the two are the same.  For example, the 
$S_3$ invariant \underline {2} $\times$ \underline {2} $\times$ \underline 
{2} $\to$ 1 is given by $(1,2) \times (1,2) \times (1,2) = 111 + 222$, 
whereas the $Z_3 \times Z_2$ analog is 
$(\omega,\omega^2) \times (\omega,\omega^2) \times (\omega,\omega^2) = 
\omega^3 + \omega^6$. It can also be checked easily that ${\cal M}_\nu$ of 
Eq.~(8) has 2 invariants and ${\cal M}_l$ of Eq.~(10) has 5 invariants in 
both cases as expected.

Diagonalizing Eq.~(12) with $V_L$ of Eq.~(19) under the condition of Eq.~(21) 
leads to the neutrino mass matrix of Eq.~(25) which yields the eigenvalues 
and eigenvectors of Eqs.~(26) to (28).  Adding the one-loop radiative 
corrections induced by the interactions of Eqs.~(30) and (31), the full 
${\cal M}_\nu$ of Eq.~(33) is then obtained.  It is very suitable for 
explaining present data on atmospheric and solar neutrino oscillations, with 
$\sin^2 2 \theta_{atm} \simeq 1$, $\tan^2 \theta_{sol} \simeq 0.4$, and 
$U_{e3}$ small but nonzero.  It also has the flexibility to allow for 
either the normal hierarchy $m_1 < m_2 < m_3$ or the inverted hierarchy 
$m_3 < m_1 < m_2$ of neutrino masses.

In this model, lepton masses come from three different sources in the Higgs 
sector: charged-lepton Dirac masses come from three Higgs doublets, neutrino 
Majorana masses come from a Higgs triplet at tree level, and the interactions 
of a charged Higgs singlet with the already present Higgs doublets at the 
one-loop level.  Singlet (right-handed) neutrinos are not needed.

To have a naturally small $\langle \xi^0 \rangle$, $m_\xi^2$ should be 
positive and large \cite{masa98}, which means that the interactions of 
$\xi$ are not observable at the electroweak scale.  On the other hand, the 
three Higgs doublets $(\phi_i^0,\phi_i^-)$ and the one Higgs singlet 
$\chi^+$ should have masses below the TeV scale, resulting in observable 
phenomena at future colliders.  The typical decay of any one of the three 
physical charged scalars is into a charged lepton and a neutrino, with large 
violations of lepton family universality.  There are also five neutral 
scalars, each decaying into a pair of charged leptons $l_i^- l_j^+$. 
Since $i \neq j$ is allowed, there should be many distinct observable 
experimental signatures.\\[5pt]

I thank M. Frigerio for discussions.  This work was supported in part by 
the U.~S.~Department of Energy under Grant No.~DE-FG03-94ER40837.

\newpage
\bibliographystyle{unsrt}

\end{document}